\documentclass[12pt]{article}
\usepackage{epsfig}
\usepackage{graphicx}
\usepackage{csquotes}
\usepackage{array}
\usepackage{float}
\usepackage{subcaption}
\usepackage{caption}
\usepackage{cleveref}
\usepackage[english]{babel}
\usepackage{setspace}
\usepackage{lineno}

\newcolumntype{P}[1]{>{\centering\arraybackslash}p{#1}}
\graphicspath{ {dpf_images/} }
\def\Title#1{\begin{center} {\Large {\bf #1} } \end{center}}

\begin{document}

\Title{Electron attenuation measurement using cosmic ray muons in the MicroBooNE LArTPC}




Varuna Meddage behalf of MicroBooNE collaberation\\
Kansas State University\\
Department of Physics\\
Manhattan,Kansas

\begin{center}
Talk presented at the APS Division of Particles and Fields Meeting (DPF 2017), July 31-August 4, 2017, Fermilab. C170731.
\end{center}

\section{Introduction}
Liquid Argon Time Projection Chamber (LArTPC) detectors provide exceptional calorimetric and track reconstruction capabilities compared to various other detector technologies. This technology has already shown great promise to be one of the ideal detector technologies to study neutrino interactions with matter. In a LArTPC, neutrinos interact with argon atoms and produce secondary charged particles (e.g protons, muons etc). These secondary charged particles ionize argon atoms and produce ionization electrons and scintillation light. The ionization electrons drift to the anode wire planes under an external electric field whereas scintillation light is collected by photo multiplier tubes (PMTs) located behind the anode wire planes. Ultimately, one can retrieve the information registered in the PMT system and anode wire planes to reconstruct 3D images of particle tracks and their energies.

One of the critical operational requirements of a LArTPC is ultra pure liquid argon. Electronegative contaminants like H\textsubscript{2}O and O\textsubscript{2} can degrade the liquid argon purity. These contaminants can capture some of the drifting ionization electrons and thus directly impact the reconstruction of particle energies. This capture process is electric field dependent where at higher electric fields ionization electrons have more chance of drifting all the way to the anode wire planes.

Equation ~\ref{eq:attenuatiion} governs how a cloud of ionization electrons gets attenuated due to the presence of electronegative contaminants inside the detector. 

\begin{equation} 
 \frac {n_e(t_{\mathrm{drift}})}{n_e(t_0)} = \exp(\frac{-t_{\mathrm{drift}}}{\tau})\label{eq:attenuatiion}
\end{equation}

Here $n_e(t_0)$ stands for the initial number of electrons whereas $n_e(t_{\mathrm{drift}})$ is the number of electrons after a time $t_{\mathrm{drift}}$. An important parameter in this equation is $\tau$, which stands for the electron lifetime. The electron lifetime contains information about the amount of electronegative contaminants present in the detector where a higher electron lifetime is indicative of low levels of contamination\cite{ICARUS1O2,ICARUS2O2}. If liquid argon is 100$\%$ pure then $\tau$ should ideally be infinite.

Figure~\ref{fig:drift1} shows the fractional loss of ionization electrons as a function of their drift distance for different electron lifetimes. It is obvious that having a higher electron lifetime in the system guarantees that more ionization electrons survive. 

\begin{figure}[htb]
\begin{center}
\includegraphics[scale=0.6]{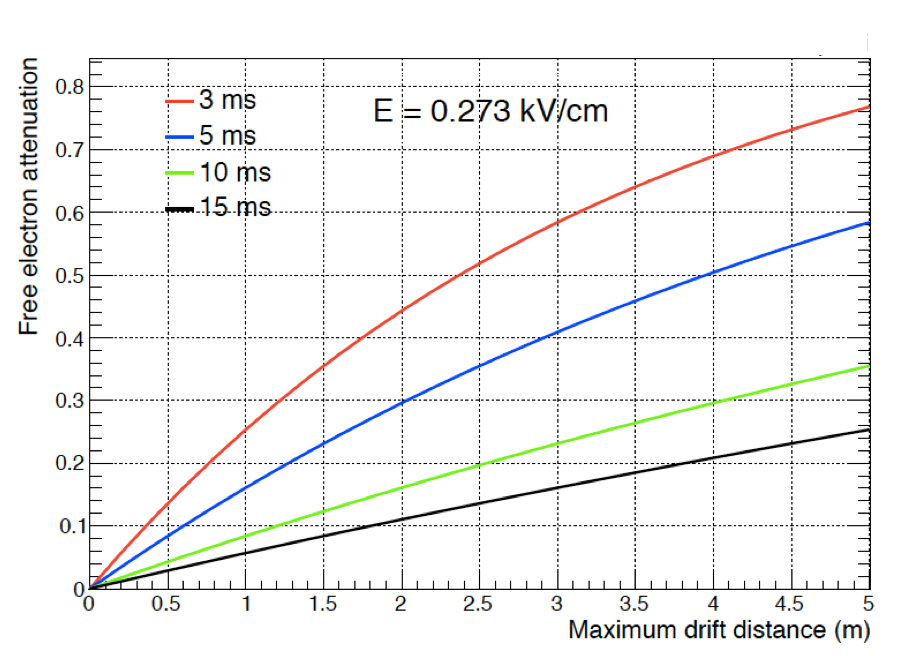}
\end{center}
\caption{Fractional loss of ionization electrons as a function of drift distance for an electric field of 0.273 kV/cm. Different colored curves correspond to different electron lifetimes ($\tau$). Graphs are created using Equation 1 for different electron lifetimes. A drift velocity of 0.1114 cm/$\mu$s (MicroBooNE's drift velocity at nominal electric field of 273 V/cm) is used to covert time into drift distance.}
\end{figure}\label{fig:drift1}

\section{The MicroBooNE LArTPC}
The MicroBooNE experiment\cite{uBDetJINST} at Fermilab uses LArTPC technology to study neutrino-argon cross sections in the 1 GeV energy regime. The other major physics goal of MicroBooNE includes addressing the low energy excess observed by the MiniBooNE experiment\cite{MiniBooNE-excess1,MiniBooNE-excess2}. MicroBooNE also serves as a critical R$\&$D step for upcoming large scale experiments like the Deep Underground Neutrino Experiment (DUNE) and the Short Baseline Neutrino (SBN) program showing the feasibility of LArTPC technology. The MicroBooNE TPC has an active volume of 85 tons of liquid argon where the cathode is kept at -70 kV thus attaining a drift electric field of 0.273 kV/cm. The anode consists of 3 wire planes (U, V and Y) where each plane has a wire pitch (separation between two of the neighboring wires, which make up the plane) of 3 mm. The two induction planes (U and V) are inclined +60\textsuperscript{o} and -60\textsuperscript{o} with respect to the vertical collection plane (Y). The maximum drift distance of the experiment is 2.56 m where it takes $\sim$2.3 ms for ionization electrons to drift from cathode to anode across the detector. The light collection system of the experiment consists of 32 8-inch PMTs located behind the anode wire planes.

The state of the art purification system in the experiment maintains the electronegative contaminants at extremely low levels which is vital for the performance of the detector. The purification system consists of two pairs of filters\cite{H20filter,O2filter} which remove H\textsubscript{2}O and O\textsubscript{2}. The initial design goal of the experiment was to maintain the O\textsubscript{2}-equivalent electronegative contaminant levels below 100 ppt to achieve an electron lifetime of 3 ms under an applied electric field of 0.5 kV/cm. The MicroBooNE detector was fully commissioned  summer of 2015 and started taking data in August 2015. By the summer of 2017, experiment collected a $\sim$$6\times10\textsuperscript{20}$ POT (protons on target) equivalent of data which contains more than $1\times10\textsuperscript{5}$ neutrino interactions to study.

\begin{figure}[htb]\label{fig:mic_dim}
\begin{minipage}[h]{0.3\textwidth}
\includegraphics[scale=0.45]{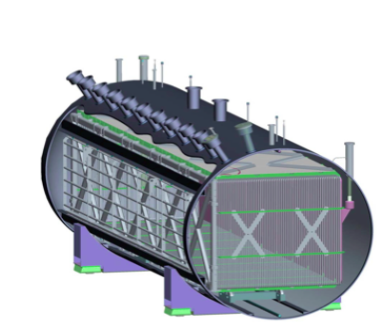}
\subcaption{\textsl{}}
\end{minipage} \hspace{0.2\textwidth}
\begin{minipage}[h]{0.3\textwidth}
\includegraphics[scale=0.45]{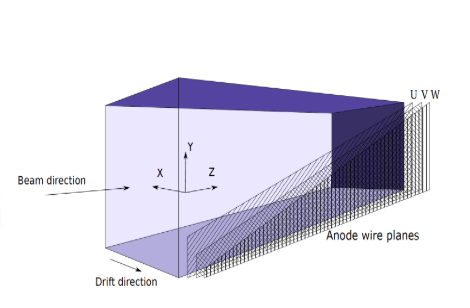}
\subcaption{\textsl{}}
\end{minipage}
\caption{(a) Layout of the MicroBooNE LArTPC. The detector is 10.4 m long (Z-direction), 2.3 m in height (Y-direction) and 2.5 m wide (X-direction). (b) Coordinate system of MicroBooNE. Two induction anode wire planes (U and V) are +60\textsuperscript{o} and -60\textsuperscript{o} with respect to the vertical whereas collection plane (Y) is vertical. }
\label{fig:mic_dim}
\end{figure}

\section{Measuring electron lifetime($\tau$) in MicroBooNE}
There are several methods one can use to measure the electron lifetime in MicroBooNE. They are:
\setlength{\parskip}{0em}
\begin{itemize}
\setlength{\parskip}{0em}
\item Gas analyzers
\setlength{\parskip}{0em}
\item Purity monitors\cite{ICARUSpm}
\setlength{\parskip}{0em}
\item Laser tracks\cite{laser1,uBlaser}
\setlength{\parskip}{0em}
\item Long minimum ionizing cosmic ray muon tracks\cite{ICARUScosmic}
\end{itemize}

Since MicroBooNE is located close to the surface, it sees abundant cosmic muons. Quantitatively, in the 4.8 ms wide readout window of MicroBooNE there are approximately 25 comic muons crossing the detector (Figure~\ref{fig:read_out}). Because of this cosmic muons provide good statistics to perform the measurement. In addition cosmic muons are uniformly distributed throughout making the measurement to be sensitive to any local variations of external electric field and electronegative contaminant levels inside the detector. But knowing the correct arrival time (or, t\textsubscript{0}) of these cosmic muons is critical for this measurement as one needs to know how long the charge drifted in the TPC. There are two types of cosmic muon datasets for which t\textsubscript{0} information is known.
\begin{itemize}
\setlength{\parskip}{0em}
\item Cosmic muons tagged by a small external cosmic ray counter\cite{MuCs}
\setlength{\parskip}{0em}
\item TPC anode-cathode crossing tracks (Crossing tracks)
\end{itemize}

Since crossing tracks have wide angular coverage compared to tracks tagged by the small external cosmic-ray counter and cover the full drift distance, these tracks are preferred. The measurement shown in this document uses the TPC anode-cathode crossing tracks.

\begin{figure}[htb]
\begin{center}
\includegraphics[scale=0.7]{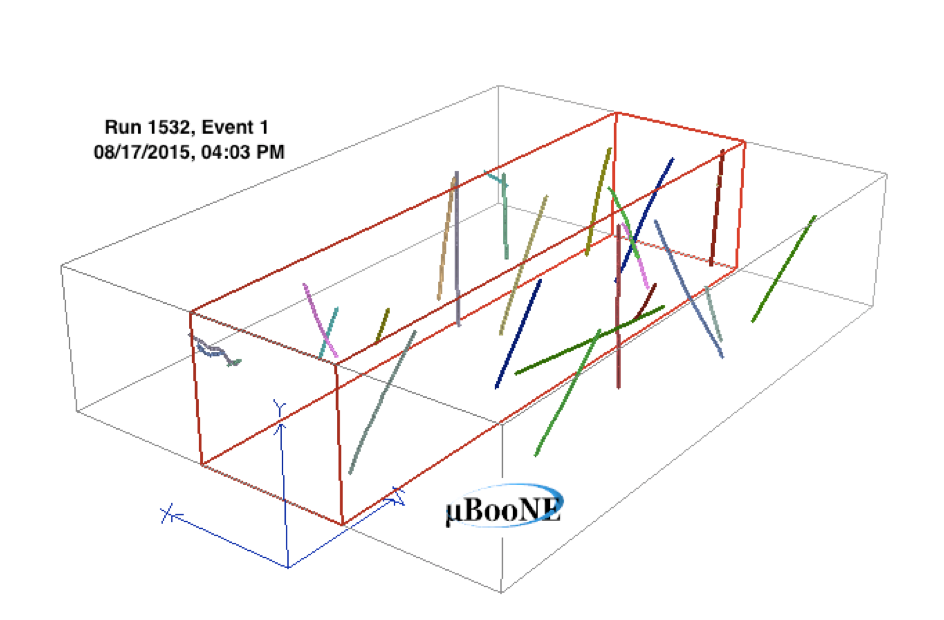}
\end{center}
\caption{Cosmic muon tracks in the MicroBooNE readout window. The three boxes show the full readout window of the MicroBooNE detector which corresponds to 4.8 ms or equivalently the total effective drift volume for a MicroBooNE readout window. The red highlighted box shows the physical volume of the TPC. The colored lines shown in the boxes are 3D reconstructed tracks. Different colors represent different tracks.}
\end{figure}\label{fig:read_out}

\section{Event Selection}
Several selection cuts are applied in the analysis to select the best quality cosmic tracks.
\begin{itemize}
\item The track projected length in the drift direction (X) must be between 250 cm to 270 cm.
\end{itemize}
Ideally the X projected track length of a crossing track should be 256 cm (drift distance of MicroBooNE). But due to track reconstruction and space charge effects a spread of this quantity is observed (Figure 4). 
\begin{itemize}
\item Exclude tracks with the angular orientation 75\textsuperscript{0}$<$$\theta_{XZ}$$<$ 105\textsuperscript{0} or 85\textsuperscript{0}$<$$\theta_{YZ}$$<$ 95\textsuperscript{0}.
\end{itemize}

$\theta_{XZ}$ is the angle defined in X-Z plane with respect to the Z direction whereas $\theta_{YZ}$ is the angle defined in the Y-Z plane with respect to the Z direction. The $\theta_{XZ}$ ($\theta_{YZ}$) cut eliminates tracks that are nearly perpendicular (parallel) to the collection plane wires. These tracks tend to be mis-reconstructed with low quality calorimetric information.

\begin{itemize}
\item[$\bullet$] The track must have a minimum of 100 hits registered in the collection wire plane.
\end{itemize}

This ensures a uniform density of hits along the drift direction and better reconstructed tracks.

\begin{itemize}
\item Exclude hits from tracks which correspond to shorted channel TPC regions.
\end{itemize}

We require that hit Y and Z coordinates are not in the regions, (-100 cm$<$Y$<$ 20 cm) and (250 cm$<$Z$<$675 cm), respectively. In the shorted channel regions of the TPC, the collection wire plane response is altered and properly reconstructed calorimetric information was not available for hits\cite{uBnoiseJINSTpre}.

\begin{figure}[htb]
\begin{center}
\includegraphics[scale=0.6]{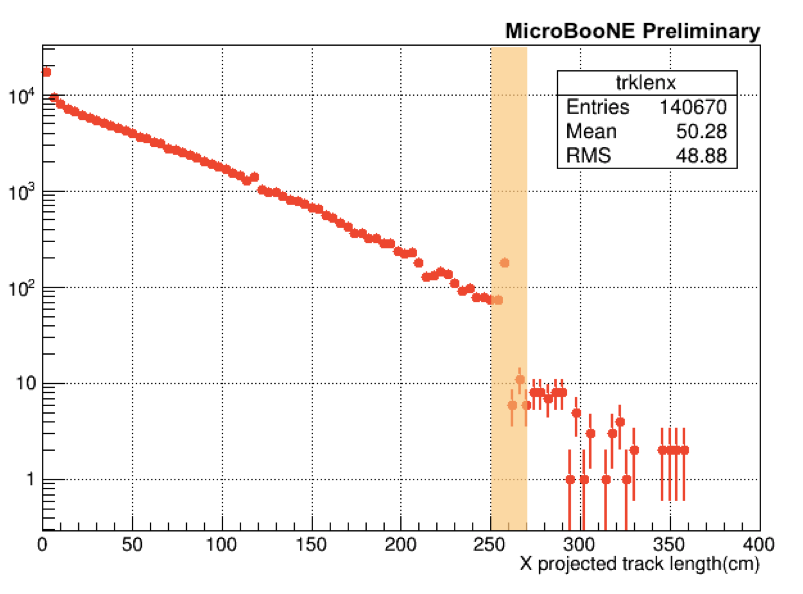}
\end{center}
\caption{Distribution of X-projected length of tracks in 10000 event sample consisting of cosmic ray data. Colored band shows the region of interest for TPC anode-cathode crossing tracks. In a $\sim$5000 event sample $\sim$2$\%$ of all tracks are crossing tracks.}
\end{figure}\label{fig:crossig_trks}

\section{Analysis Method}
The following method is used to calculate the electron attenuation (electron lifetime) from a set of crossing tracks which satisfy all the selection criterion described in the previous section.

\begin{itemize}
\item The start time (t\textsubscript{0}) of the track is calculated : For TPC crossing tracks either light information or hit information can be used to extract t\textsubscript{0}. At the time of this analysis, light reconstruction wasn't reliable. So hit information is used to get t\textsubscript{0}. Minimum drift coordinate of the crossing track provides the t\textsubscript{0}. Once t\textsubscript{0} is known, each drift coordinate of the track is corrected using Equation~\ref{eq:x_correction}.
\end{itemize}


\begin{equation}
X_{\mathrm{corrected}} = X_{\mathrm{reconstructed}}-X_{\mathrm{minimum}}\label{eq:x_correction}
\end{equation}

\begin{itemize}
\item The full drift time window (2.2 ms) is split into smaller 100 $\mu$s wide time windows (Figure 5(a)).
\end{itemize}

\begin{itemize}
\item For each drift time bin, the charge deposited per unit length (dQ/dx) distribution is produced and fit with a Landau convoluted with a Gaussian function\cite{lgfit} to get the Most Probable Value (MPV) of dQ/dx representing that time bin (Figure 6).
\end{itemize}

\begin{itemize}
\item[$\bullet$] All 22 most probable dQ/dx values are plotted against drift time and this final distribution is fit with a function f(t) to get the final Q\textsubscript{A}/Q\textsubscript{C} charge ratio (Figure 5(b)).
\end{itemize}

The final Q\textsubscript{A}/Q\textsubscript{C} charge ratio is calculated using Equation~\ref{eq:qa_qc} which gives the fractional change in charge due to capture by electronegative contaminants when a cloud of ionization electrons drift from cathode to anode. Ideally f(t) should be exponential. But the presence of space charge and other effects can skew the dQ/dx distribution resulting in non-exponential shapes. Second-order polynomial and exponential plus constant functions are also used to fit to the final distribution. Once the Q\textsubscript{A}/Q\textsubscript{C} charge ratio is calculated, assuming an exponential behavior, one can translate it into an electron lifetime using Equation~\ref{eq:charge_to_elife}. It was observed that for most of the runs, the final Q\textsubscript{A}/Q\textsubscript{C} value is greater than 1 due to space charge effects.

\begin{equation}
\frac{Q_A}{Q_C} = \frac{f(2200~\mu s)}{f(0~\mu s)}\label{eq:qa_qc}
\end{equation}

In Equation 3, the numerator and denominator stand for the amount of charge arriving at the anode after 2.2 ms and charge leaving the cathode at 0 ms, respectively.

\begin{equation}
\frac{Q_A}{Q_C} = \exp(\frac{-t_{\mathrm{drift}}}{\tau})\label{eq:charge_to_elife}
\end{equation}

Here $t_{\mathrm{drift}}$ should be replaced by 2200 $\mu$s to get $\tau$.

\begin{figure}[htb]
\begin{minipage}[h]{0.3\textwidth}
\includegraphics[width=7cm, height=5cm]{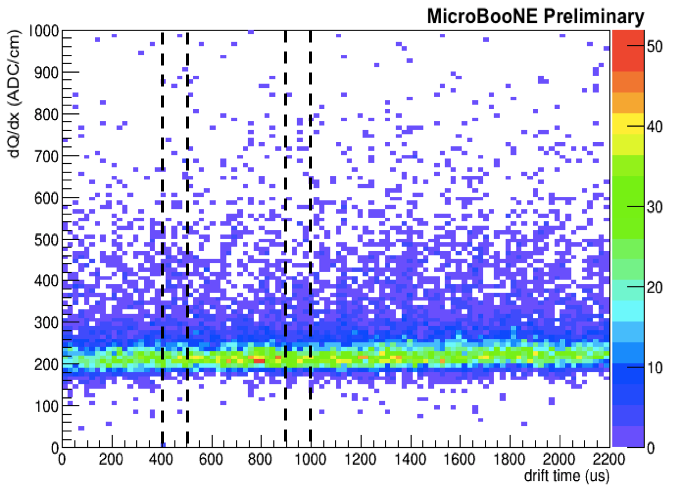}
\subcaption{\textsl{}}
\end{minipage} \hspace{0.2\textwidth}
\begin{minipage}[h]{0.3\textwidth}
\includegraphics[width=7cm, height=5cm]{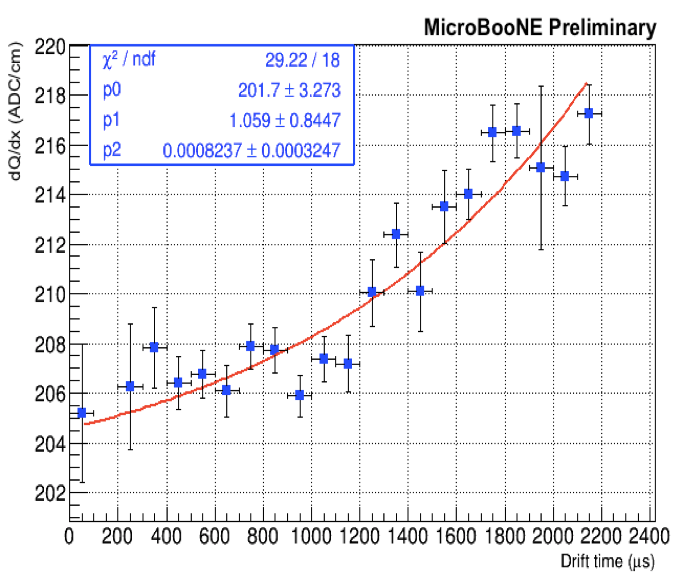}
\subcaption{\textsl{}}
\end{minipage}
\caption{(a) dQ/dx vs drift time scatter plot. The full drift time window (2.2 ms) is split into 100 $\mu$s wide time bins. Regions between the dashed vertical lines represent smaller time bins. (b) Distribution of most probable dQ/dx values (from each smaller time bin) as  a function of drift time. Distribution is fit with a second-order polynomial to get the final Q\textsubscript{A}/Q\textsubscript{A} charge ratio.}
\label{fig:splitting_final_dqdx}
\end{figure}

\begin{figure}[htb]
\begin{minipage}[h]{0.3\textwidth}
\includegraphics[width=7cm, height=5cm]{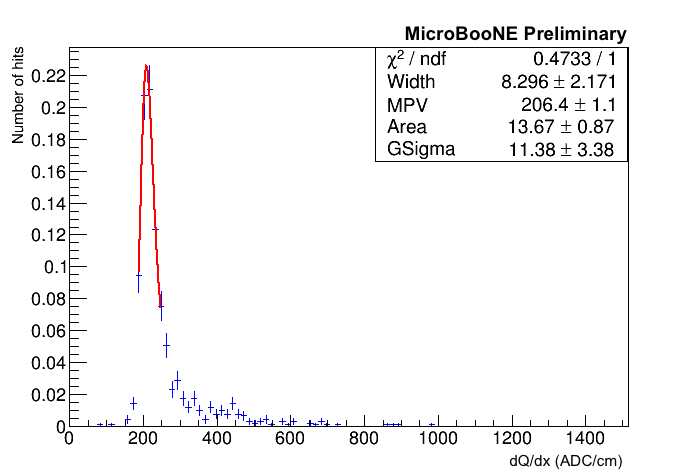}
\subcaption{\textsl{}}
\end{minipage} \hspace{0.2\textwidth}
\begin{minipage}[h]{0.3\textwidth}
\includegraphics[width=7cm, height=5cm]{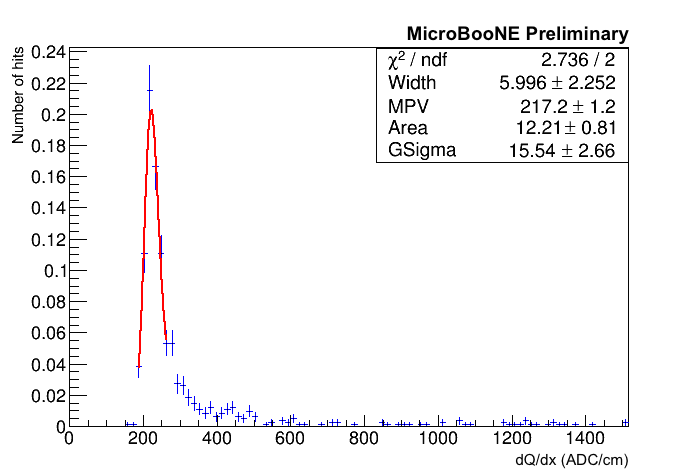}
\subcaption{\textsl{}}
\end{minipage}
\caption{(a) dQ/dx distribution produced for the time bin 400 $\mu$s to 500 $\mu$s. (b) dQ/dx distribution produced for the time bin 2100 $\mu$s to 2200 $\mu$s. Distributions are fit with a Landau convoluted wih a Gaussian to extract the most probable dQ/dx value in that time range.}
\label{fig:small_dqdx_dis}
\end{figure}

\section{Space Charge Effects}
Since MicroBooNE is surface-based, a lot of cosmic activity is present in the detector. As a result of slow moving positive argon ions get accumulated leading to distortions in the applied electric field both in its magnitude (Figure 7(a)) and direction. Once the directionality of the electric field is impacted, it can affect reconstructed position of ionization electron clusters (Figure 7(b)) and calorimetric information (dQ/dx). The magnitude change of the electric field directly affects electron-ion recombination. The lower the applied electric field the stronger the recombination gets. From MicroBooNE simulations, it is estimated that the electric field close to the cathode increases by $\sim$12$\%$ whereas at the anode it decreases by $\sim$5$\%$. The modified box model is used to describe the electron-ion recombination\cite{recomb} in MicroBooNE where it shows dQ/dx values increasing by $\sim$3.55$\%$ close to the cathode and decreasing by $\sim$1.2$\%$ close the to anode. (See section 9.3 for more details on recombination). \\


In order to see the impact of space charge effects on the final measurement, a 3D space charge model is implemented in simulation and isotropic single muon monte carlo samples are created with a very high electron lifetime switching off all other effects such as diffusion which can affect the attenuation mesurement. The positive slope of Figure 8(b) shows final Q\textsubscript{A}/Q\textsubscript{A} charge ratio can be greater than 1 due to space charge effects.

\begin{figure}[htb]
\begin{minipage}[h]{0.3\textwidth}
\includegraphics[scale=0.6]{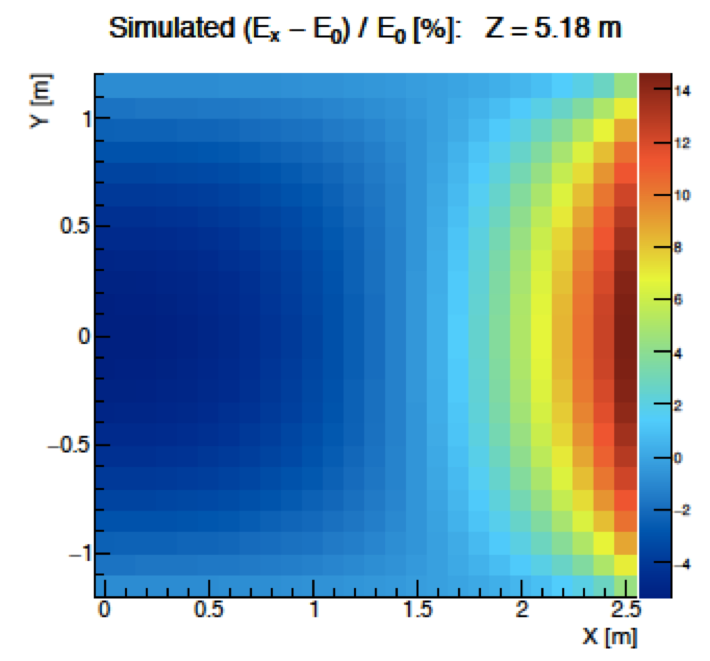}
\subcaption{\textsl{}}
\end{minipage} \hspace{0.2\textwidth}
\begin{minipage}[h]{0.3\textwidth}
\includegraphics[scale=0.6]{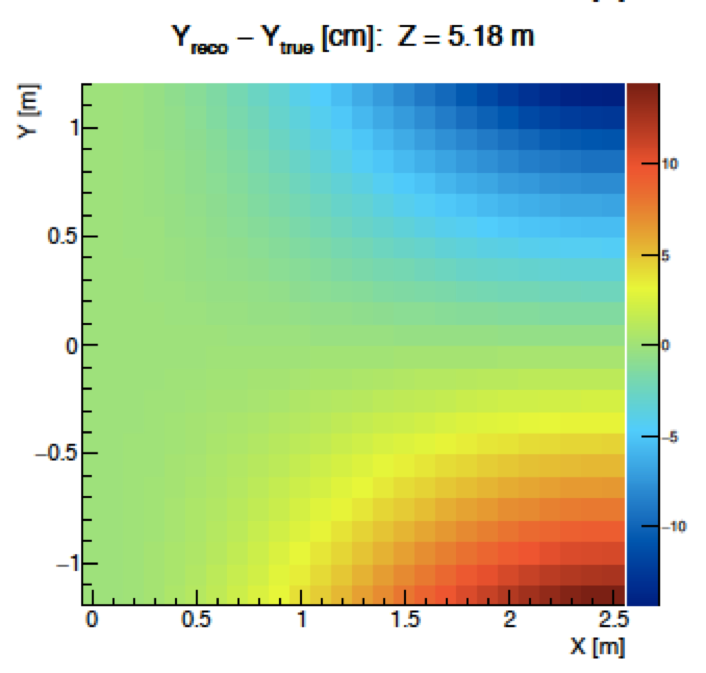}
\subcaption{\textsl{}}
\end{minipage}
\caption{(a) Illustration of simulated effect of space charge on the X component of the electric field as a function of the X and Y coordinates.. (b) Simulated effect of space charge on the reconstructed Y position of an electron cluster as a function of the X and Y coordinates. Both of these plots are created for the central Z region where space charge effects are thought to be maximal.}
\label{fig:mag_dir_simulate}
\end{figure}

\begin{figure}[htb]
\begin{minipage}[h]{0.3\textwidth}
\includegraphics[width=7cm, height=5cm]{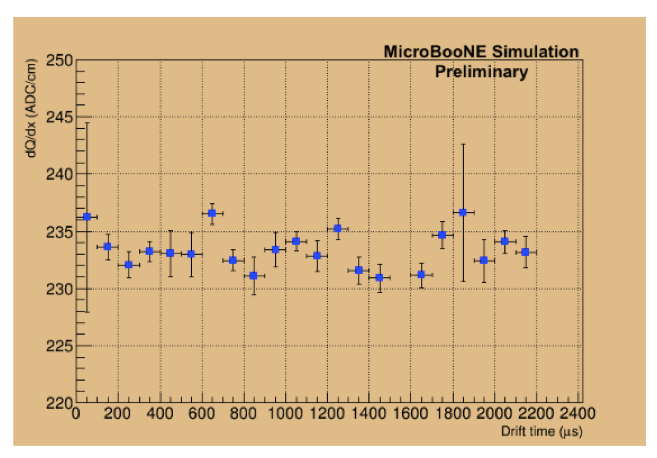}
\subcaption{\textsl{}}
\end{minipage} \hspace{0.2\textwidth}
\begin{minipage}[h]{0.3\textwidth}
\includegraphics[width=7cm, height=5cm]{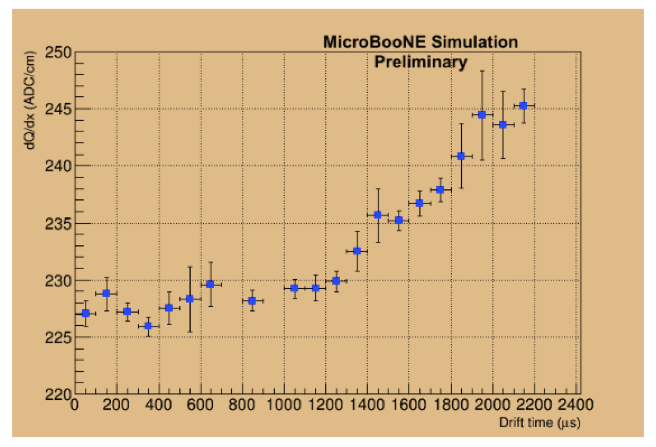}
\subcaption{\textsl{}}
\end{minipage}
\caption{(a) dQ/dx distribution of the monte carlo sample with no space charge effects. (b) dQ/dx distribution of the monte carlo sample with space charge effects including distortions in both the direction and magnitude of the electric field.}
\label{fig:sp_mc}
\end{figure}

\subsection{Space Charge Correction}
As space charge effects (SCE) significantly impact the final measurement, a correction is derived using the following procedure. 

\begin{description}
\item[$\bullet$] The correction \enquote{C} is derived for dQ/dx values.
\end{description}

The correction for dQ/dx values is defined for all 22 drift bins using Equation 5 based on the two dQ/dx distributions in Figure 8.

\begin{equation}
C = \frac {(\frac {dQ}{dx})_{\mathrm{SCE=ON}} - (\frac {dQ}{dx})_{\mathrm{SCE=OFF}}}   {(\frac {dQ}{dx})_{\mathrm{SCE=ON}}} \label{eq:cor_1}
\end{equation}

\begin{description}
\item[$\bullet$] A third order polynomial fit is used to extract corrections.
\end{description}

To make the derived corrections uniform, the correction \enquote{C} is plotted as a function of drift time and the resultant distribution is fit with a third order polynomial (Figure 9). The final corrections \enquote{$\tilde{\mathrm{C}}$} for each bin are extracted using the polynomial fit. This space charge correction can be applied to data using Equation~\ref{eq:cor_dqdx}.

\begin{equation}\label{eq:cor_dqdx}
(\frac{dQ}{dx})_{\mathrm{corrected}} = \frac{dQ}{dx}(1-\tilde{C})\label{eq:cor_dqdx}
\end{equation}

Here, dQ/dx stands for the uncorrected value from data including space charge effects. 


\begin{figure}[htb]
\begin{center}
\includegraphics[scale=0.4]{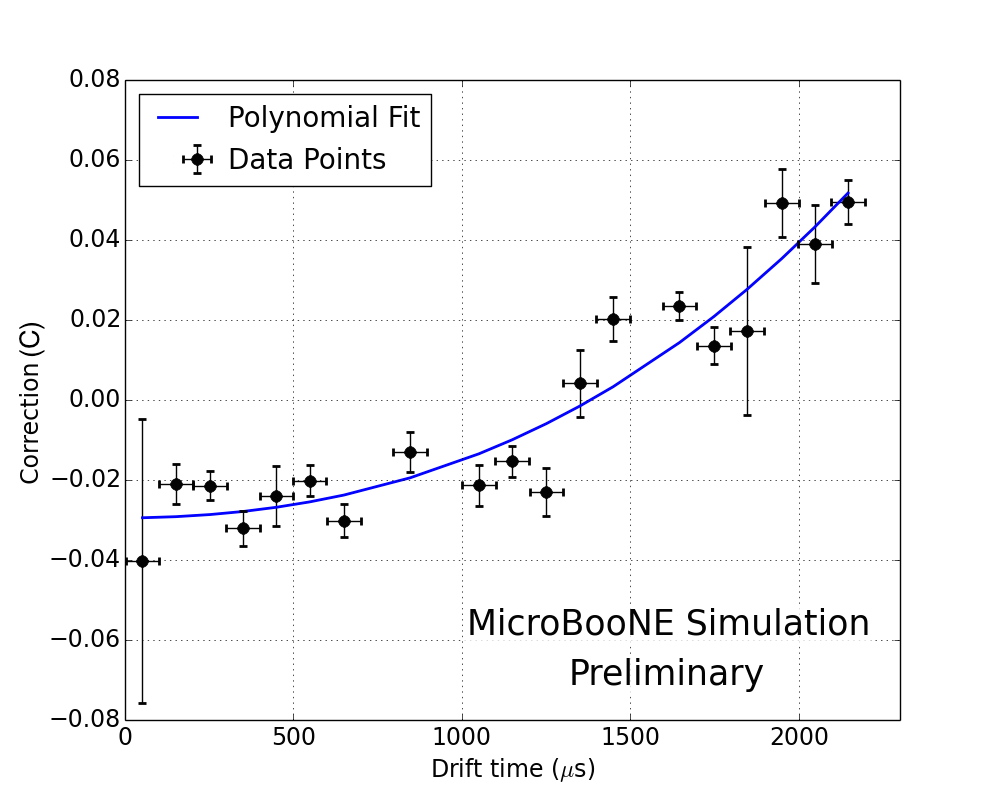}
\end{center}
\caption{Plot of space charge corrections versus drift time.  A third order polynomial is used to fit the points to extract corrections for all 22 bins.}
\end{figure}\label{fig:pol3_fit_11}

\section{Data Sample}

Cosmic muon data ranging from 02/16/2016 to 04/21/2016 is used in this analysis. Each dataset has approximately 5000 events processed to study the variation of electron attenuation on a daily basis. The time difference between two consecutive data sets is $\sim$24 hours and the detector spans a range of low purity conditions to high purity conditions during the selected time period. Some of the datasets are missing in this time period due to data processing problems.

\section{Variation of Q\textsubscript{A}/Q\textsubscript{C}}

The plot in the Figure 10 shows the variation of the Q\textsubscript{A}/Q\textsubscript{C} charge ratio over time by analyzing 56 datasets described in section 7 using the procedure explained in section 5 with and without the space charge corrections derived in section 6.1.

\begin{figure}[htb]
\begin{center}
\includegraphics[scale=0.6]{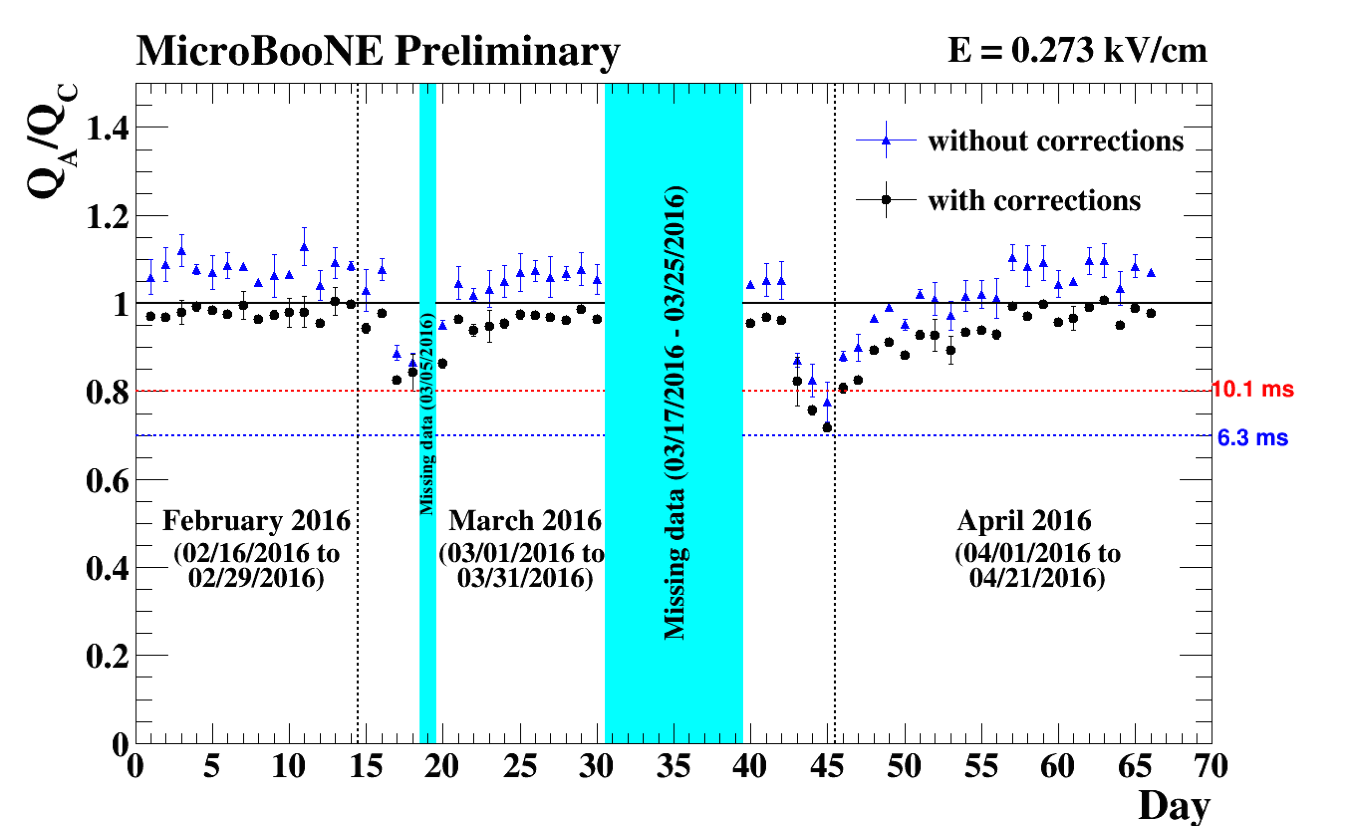}
\end{center}
\caption{Variation of Q\textsubscript{A}/Q\textsubscript{C} over time with and without space charge corrections. Colored bands show the regions of missing data due to data processing problems.}
\end{figure}\label{fig:inivar}

\section{Systematic Uncertainties}
A number of possible sources of systematic uncertainty are considered in this analysis. Among them, the dominant systematic uncertainty comes from the space charge correction
 used in the analysis. The other systematics that effect this measurement in a minor way are recombination and diffusion. Both data-drive and monte-carlo-based techniques are used in addressing systematics. 

\subsection{Space charge correction uncertainty}
Space charge corrections described in section 6.1 are derived from the 3D space charge model implemented in the simulations which is only valid for the central Z region. So to account for space charge model dependency, a large uncertainty is associated to the final Q\textsubscript{A}/Q\textsubscript{C} charge ratio values corresponding to 50$\%$ of the difference of the two Q\textsubscript{A}/Q\textsubscript{C} charge ratios before and after the space charge correction. This is done separately for all 56 data sets where the systematic varies between 1.4$\%$ to 7.5$\%$  with an average uncertainty of 4.6$\%$. The systematic is assigned to be 5.0$\%$ of the final Q\textsubscript{A}/Q\textsubscript{C} charge ratio for all data sets.

\subsection{Diffusion}
Diffusion can smear the ionization electron cloud as they drift from cathode to anode by affecting the charge collected by the anode plane wires. In particular, the transverse component of the diffusion which is perpendicular to the drift direction can leak the charge to neighboring wires from the target wire inducing modifications to the reconstructed dQ/dx values. Any modifications in dQ/dx values directly impacts the measured  Q\textsubscript{A}/Q\textsubscript{C} charge ratio. To calculate the systematic uncertainty introduced by diffusion, two monte carlo single isotropic muon samples are used, where in one both the longitudinal and transverse diffusion components are turned on while in the other the diffusion effects are turned off. The systematic is assigned to be the difference between  Q\textsubscript{A}/Q\textsubscript{C} charge ratios with and without diffusion, which is 2.0$\%$ of the final value.


\subsection{Recombination}
As explained before, in the current MicroBooNE simulation the electron-ion recombination is described by modified box model. Equation~\ref{eq:recombination_mbm} shows the relationship between the recombination factor \enquote{$R_{box}$} and the electric field.

\begin{equation}
R_{box} = \frac {ln(\alpha + \frac {\beta_p} {\rho\varepsilon}\cdot\frac{dE}{dx} )}  { \frac {\beta_p} {\rho\varepsilon}\cdot\frac{dE}{dx}}\label{eq:recombination_mbm}
\end{equation}

Here $\varepsilon$ is the electric field of MicroBooNE (0.273 kV/cm), $\rho$ is liquid argon density (1.38 g/cm\textsuperscript{3} at a pressure 18.0 psia), $\beta$\textsubscript{p}=0.212 +/- 0.002 (kV/cm)(g/cm\textsuperscript{2})/MeV and $\alpha$=0.93+/-0.02. The values for $\beta$\textsubscript{p} and $\alpha$ were calculated by Argoneut experiment which had an operating electric field of 0.481 kV/cm\cite{recomb}.\\

To account for the model dependency of recombination in the electron attenuation measurement, two single isotropic muon monte carlo samples are generated, one having the default values for $\beta$\textsubscript{p} and $\alpha$ in the recombination model whereas in the other these two values are maximally changed ( by 0.01 (kV/cm)(g/cm\textsuperscript{2})/MeV and 0.1 respectively). In both of these samples all the other effects are turned off such as diffusion with the exception the change in the magnitude of the electric field due to space charge effects. The percentage difference of the Q\textsubscript{A}/Q\textsubscript{C} charge ratios between samples with modified parameter and default parameter settings is found to be 1.0$\%$ with respect to the Q\textsubscript{A}/Q\textsubscript{C} charge ratio extracted for default parameter setting. This is assigned to be the systematic value coming from recombination model dependency. Table 1 summarizes all the systematics considered in the analysis.

\begin{table}[b]\label{table:sys_summary}
\centering
\begin{tabular}{ l|ccc}
\hline
 Systematic & Uncertainity ($\%$)\\
 \hline
 Space charge correction   & 5.0\\
 Recombination & 1.0\\
 Diffusion & 2.0\\
\hline
Total & 5.5\\
 \hline
\end{tabular}
\caption{Systematic uncertainties in the final Q\textsubscript{A}/Q\textsubscript{C} charge ratio. The total is calculated by adding individual systematics in quadrature.}
\end{table}


\section{Results}

Figure 11 shows the  variation of the Q\textsubscript{A}/Q\textsubscript{C} charge ratio over time in MicroBooNE with both statistical and systematic uncertainties folded in and the space charge correction. It can be observed that the Q\textsubscript{A}/Q\textsubscript{C} charge ratio is very high even taking into account the systematic uncertainties. During stable purity conditions, (excluding the two dip regions in the figure) the Q\textsubscript{A}/Q\textsubscript{C} value changes between 0.88+/-0.04 and 1.01+/-0.05. Using Equation~\ref{eq:charge_to_elife} the lowest corresponding electron lifetime in this period can be found to be 18 ms which corresponds to a maximum charge loss of 12$\%$ and an O\textsubscript{2} equivalent contamination level of 17 ppt. The lowest Q\textsubscript{A}/Q\textsubscript{C} value of the entire time period studied is 0.72+/-0.03 which corresponds to a maximum charge loss of 28$\%$, an electron lifetime of 6.8 ms, and an O\textsubscript{2} equivalent contamination level of 44 ppt.

\begin{figure}[htb]\label{fig:final_var}
\begin{center}
\includegraphics[scale=0.7]{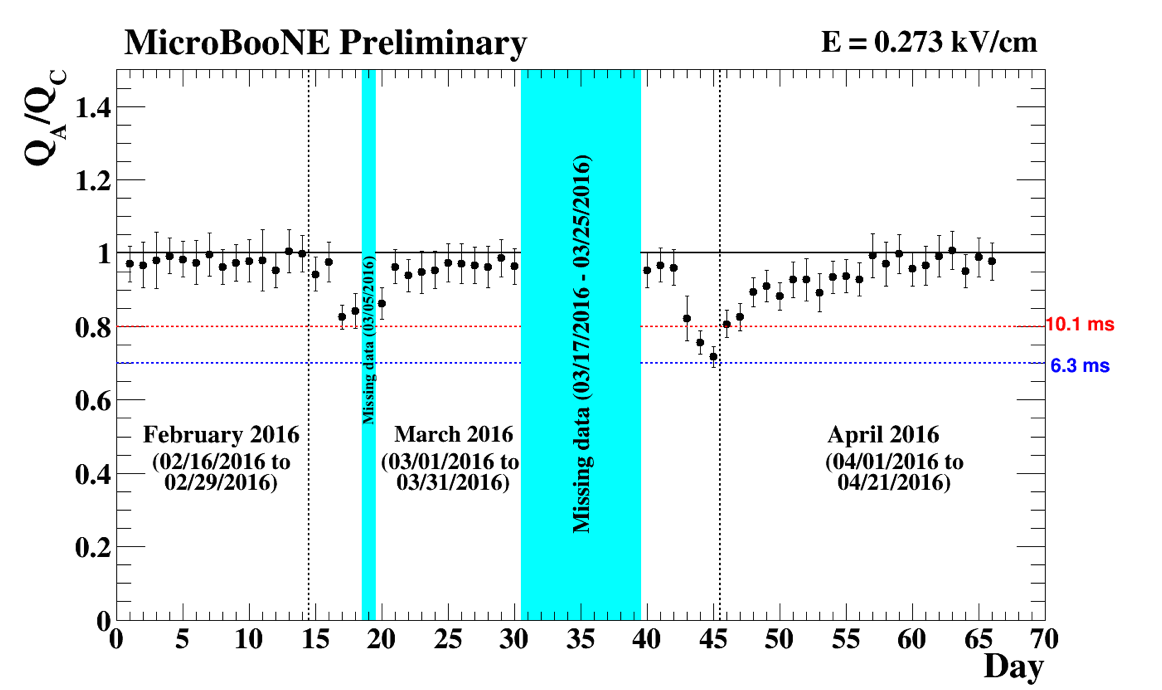}
\end{center}
\caption{Variation of  Q\textsubscript{A}/Q\textsubscript{C} charge ratio over time with both statistical and systematic uncertainties and space charge corrections. Colored bands show the region where missing data lies.}
\end{figure}

\section{Summary and Conclusions}

This is of the first measurement of the liquid argon purity in MicroBooNE expressed in terms of Q\textsubscript{A}/Q\textsubscript{C} charge ratio (ratio of collected charge to deposited charge) using cosmic ray muons. During the time period of 02/16/2016 - 04/21/2016, Q\textsubscript{A}/Q\textsubscript{C} values range from 0.72+/-0.03 to 1.01+/-0.05 indicating a very high liquid argon purity inside the detector. The lowest Q\textsubscript{A}/Q\textsubscript{C} charge ratio recorded in the entire duration is 0.72+/-0.03 which corresponds to an electron lifetime of 6.8 ms and an O\textsubscript{2} equivalent contamination level of 44 ppt resulting in a charge loss of 28$\%$ for the full drift path. Systematic uncertainties dominate over statistical fluctuations in the entire time period studied. During stable purity conditions the lowest Q\textsubscript{A}/Q\textsubscript{C} charge ratio observed is 0.88+/-0.04 with a corresponding electron lifetime of 18 ms and an O\textsubscript{2} equivalent contamination level of 17 ppt. These results are indicative of MicroBooNE having a very high electron lifetime which clearly exceeds the early design goal of 3 ms.


\begin{thebibliography}{99}

\bibitem{ICARUS1O2}
E. Buckley \textit{et al.,} ``A study of ionization electrons drifting over large distances in liquid argon'', Nucl. Instrum. and Methods in Physics Research A 275, 364-372 (1989). 

\bibitem{ICARUS2O2}
A. Bettini \textit{et al.,} ``A study of the factors affecting the electron lifetime in ultra-pure liquid argon'', Nucl. Instrum. and Methods in Physics Research A 305, 177-186 (1991).

\bibitem{ICARUSpm}
S. Amoruso \textit{et al.,} ``Analysis of the liquid argon purity in the ICARUS T600 TPC'', Nucl. Instrum. and Methods in Physics Research A 516, 68-79 (2014).


\bibitem{ICARUScosmic}
M. Antonello \textit{et al.,} ``Experimental observation of an extremely high electron lifetime with the ICARUS-T600 LAr-TPC'', JINST 9, P12006 (2014).

\bibitem{laser1}
A. Ereditato \textit{et al.,} ``Design and operation of ARGONTUBE: a 5~m long drift liquid argon TPC'', JINST 8, P07002 (2013).


\bibitem{uBDetJINST}
The MicroBooNE Collaboration, ``Design and construction of the MicroBooNE detector'', JINST 12, P02017 (2017).

\bibitem{MiniBooNE-excess1}
A.~A.~Aguilar-Arevalo \textit{et al.,}
  ``Unexplained Excess of Electron-like Events from a 1~GeV Neutrino Beam'',
  Phys. Rev. Lett. {\bf102}, 101802 (2009).

\bibitem{MiniBooNE-excess2}
  A.~A.~Aguilar-Arevalo \textit{et al.,}
  ``A Combined $\nu_{\mu}\rightarrow \nu_e$ and $\bar \nu_{\mu} \rightarrow \bar \nu_e$ Oscillation Analysis of the MiniBooNE Excesses'',
  Phys. Rev. Lett. {\bf110}, 161801 (2013).

\bibitem{H20filter}
Sigma-Aldrich, P.O. Box 14508, St. Louis, MO 63178 USA.

\bibitem{O2filter}
BASF Corp., 100 Park Avenue, Florham Park, NJ 07932 USA.

\bibitem{larsoft}
E. D. Church, ``Larsoft: A software package for liquid argon time projection drift chambers'', arXiv:1311.6774 (2016).


\bibitem{caloArgoneut}
The ArgoNeuT Collaboration, ``Analysis of a large sample of neutrino-induced muons with the ArgoNeuT detector'', JINST 7, P10020 (2012).

\bibitem{uBnoiseJINSTpre}
The MicroBooNE Collaboration, "Noise Characterization and Filtering in the MicroBooNE Liquid Argon TPC", arXiv:1705.07341, JINST 12, P08003 (2017).

\bibitem{recomb}
The ArgoNeuT Collaboration, ``A study of electron recombination using highly ionizing particles in the ArgoNeuT Liquid Argon TPC'', JINST 8, P08005 (2013). 

\bibitem{lgfit}
Convoluted Landau and Gaussian Fitting Function, \newline
https://root.cern.ch/root/html/tutorials/fit/langaus.C.html

\bibitem{longbo}
The LongBo Collaboration, ``Design and operation of LongBo: a 2~m long drift liquid argon TPC'', JINST 10, P07015 (2015).


\bibitem{35ton}
The DUNE 35-ton Prototype Collaboration, ``Evidence of Impurity Stratification in the DUNE 35 ton Prototype Cryostat'', FERMILAB-TM-2642-ND (2017).

\bibitem{icarusloc}
The ICARUS Collaboration, ``Measurement of the $\mu$ decay spectrum with the ICARUS
liquid Argon TPC'', Eur. Phys. J. C 33, 233 (2004).

\bibitem{uBlaser}
A. Ereditato \textit{et al.,} ``A steerable UV laser system for the calibration of liquid argon time projection chambers'', JINST 9, T11007 (2014).

\bibitem{MuCs}
MicroBooNE collaboration, "Measurement of Cosmic Ray Reconstruction Efficiencies in the MicroBooNE LAr TPC Using a Small External Cosmic Ray Counter", arXiv:1707.09903, submitted to JINST

























\end{thebibliography}
\end{document}